**Current-induced spin-orbit field in permalloy interfaced with ultrathin Ti and Cu**

Ryan W. Greening[1], David A. Smith[1], Youngmin Lim[1], Zijian Jiang[1], Jesse Barber[1], Steven Dail[1,2], Jean J. Heremans[1], Satoru Emori[1,*]

1. Department of Physics, Virginia Tech, Blacksburg, VA 24061

2. Academy of Integrated Science, Virginia Tech, Blacksburg, VA 24061

* email: semori@vt.edu

How spin-orbit torques emerge from materials with weak spin-orbit coupling (e.g., light metals) is an open question in spintronics. Here, we report on a field-like spin-orbit torque (i.e., in-plane spin-orbit field transverse to the current axis) in $SiO_2$-sandwiched permalloy (Py), with the top Py-$SiO_2$ interface incorporating ultrathin Ti or Cu. In both $SiO_2$/Py/Ti/$SiO_2$ and $SiO_2$/Py/Cu/$SiO_2$, this spin-orbit field opposes the classical Oersted field. While the magnitude of the spin-orbit field is at least a factor of 3 greater than the Oersted field, we do not observe evidence for a significant damping-like torque in $SiO_2$/Py/Ti/$SiO_2$ or $SiO_2$/Py/Cu/$SiO_2$. Our findings point to contributions from a Rashba-Edelstein effect or spin-orbit precession at the (Ti, Cu)-inserted interface.

An electric current in a material with spin-orbit coupling generally gives rise to a non-equilibrium spin accumulation [1–6], which can then exert torques – i.e., spin-orbit torques (SOTs) – on magnetization in an adjacent magnetic medium [7–9]. SOTs are often classified into two symmetries: damping-like SOT that either counters or enhances magnetic relaxation, and field-like SOT (or "spin-orbit field") that acts similarly to a magnetic field. Next generations of nanomagnetic computing devices may benefit from an improved understanding of mechanisms for SOTs and the discovery of new thin-film systems enabling large SOTs.

While most efforts have focused on conductors known for strong spin-orbit coupling (e.g., 5d transition metals, topological insulators, etc.) [7,8], recent reports have shown SOTs in ferromagnets interfaced with materials that are not expected to exhibit significant spin-orbit coupling [10–14]. For example, a large damping-like SOT has been reported in ferromagnetic $Ni_{80}Fe_{20}$ (permalloy, Py) interfaced with partially oxidized Cu [10,11]; quantum-interference transport measurements have revealed that Cu with an oxidation gradient can, in fact, exhibit enhanced spin-orbit coupling comparable to that in heavier metals (e.g., Au) [15]. As another example of SOTs that emerge by incorporating seemingly weak spin-orbit materials, Py interfaced with a Ti seed layer and $Al_2O_3$ capping layer exhibits a sizable field-like SOT [12]. The key observed features of this spin-orbit field in Ti/Py/$Al_2O_3$ [12] are: (1) it points in-plane and transverse to the current axis, irrespective of the magnetization orientation in Py; (2) its magnitude scales inversely with the Py thickness, i.e., it is interfacial in origin; (3) it is modified significantly by the addition of an insertion layer (e.g., Cu) at the Py-$Al_2O_3$ interface. Ref. [12] claims that this spin-orbit field is governed by a Rashba-Edelstein effect (REE) [1,5,16,17] at the Py/$Al_2O_3$ and Cu/$Al_2O_3$ interfaces. However, the complicated stack structures of $SiO_2$(substrate)/Ti/Py/(Cu/)$Al_2O_3$ with multiple dissimilar interfaces in Ref. [12] obscure the mechanisms of the spin-orbit field, particularly the roles played by the Ti and Cu layers.

Here, by using simpler stack structures, we gain insight into the impact of ultrathin Ti and Cu interfacial insertion layers on the current-induced spin-orbit field in Py at room temperature. Specifically, we have characterized the total current-induced transverse field $H_{I,tot}$ in $SiO_2$/Py/Ti/$SiO_2$ (Py/Ti) and $SiO_2$/Py/Cu/$SiO_2$ (Py/Cu) with the second-order planar Hall effect (PHE) [18,19] and spin-torque ferromagnetic resonance (ST-FMR) [20]. From the observed $H_{I,tot}$ and estimated classical Oersted field $H_{Oe}$ in each stack structure, we extract the spin-orbit field $H_{so}$ via

$$H_{so} = H_{I,tot} - H_{Oe}. \quad (1)$$

We find that Py/Ti and Py/Cu exhibit $H_{so}$ that opposes $H_{Oe}$ with a similar magnitude, i.e., at least 3 times greater than $H_{Oe}$. While this field-like SOT is well above our detection limit, we observe no evidence for a significant damping-like SOT in Py/Ti or Py/Cu. We deduce that the Rashba field at the (Ti, Cu)-inserted interface plays a key role in the observed $H_{so}$.

We patterned Py/Ti and Py/Cu, along with a control symmetric stack of $SiO_2$/Py/$SiO_2$ (sym-Py), by photolithography and liftoff into Hall crosses (for second-order PHE measurements) and rectangular microstrips (for ST-FMR measurements). The substrate was Si (001) covered with 50-nm-thick thermally grown oxide. We used rf-sputtered $SiO_2$ as both the buffer and capping layers to preserve the structural symmetry of the sym-Py control stack. The metallic Py, Ti, and Cu layers were deposited by dc sputtering. The nominal deposited layer thicknesses were 3 nm for $SiO_2$, 3 nm for Py, and 0.5 nm for Ti and Cu. Static magnetic properties of the sym-Py, Py/Ti, and Py/Cu films are summarized in the Supplementary Material. The patterned Hall crosses were 100 and 200 μm wide, with essentially identical results obtained for both device widths, whereas the ST-FMR microstrips had widths of 50 μm. Both device types were contacted by thermally evaporated Cr (3 nm)/Au (100 nm) electrodes, patterned with an additional layer of photolithography and liftoff.

By four-point measurements on double Hall crosses, we obtained the sheet resistance for each film stack structure: 320 Ohm/sq for sym-Py, 250 Ohm/sq for Py/Ti, and 200 Ohm/sq for Py/Cu. The smaller resistance values for Py/Ti and Py/Cu, compared to sym-Py, suggest that ultrathin Ti and Cu produce an additional conductive path. The conductance of the Py layer in Py/Ti and Py/Cu may also be higher than in sym-Py, due to the Ti and Cu insertion layers protecting the top Py surface from oxidation. Both scenarios result in the top portions of the Py/Ti and Py/Cu stacks contributing more to conductance than the bottom portions with the direct $SiO_2$-Py interfaces. We can therefore determine the direction of the Oersted field $H_{Oe}$ acting on the magnetization in Py; referring to Fig. 1(a) with the Py/Cu stack as an example, with a conventional (positive) charge current along the +$x$ direction, a higher current density in the top portion of the stack generates a net $H_{Oe}$ along the +$y$ direction within the Py layer.

To quantify the distribution of in-plane current density, for simplicity, we treat the Ti (or Cu) and Py layers as parallel resistors and fix the resistance of Py to that found from sym-Py. We estimate the fraction of the current in Ti (Cu) to be $f_{Ti} \approx 20\%$ ($f_{Cu} \approx 40\%$). This approximation likely overestimates the current in Ti and Cu, since the Py layer in Py/Ti and Py/Cu may be more conductive than that in sym-Py. Nevertheless, this approximation yields a useful upper bound of $H_{Oe}$ in the stack structures via $|H_{Oe}| = |I_{dc}|f_{(Ti,Cu)}/(2w)$, where $I_{dc}$ represents the total in-plane current through the device and $w$ the device width.

In addition to the sym-Py, Py/Ti, and Py/Cu stacks, we also used Hall crosses and microstrips of Ta(3)/Py(2.5)/Pt(4) from a prior study [21] as an additional control sample to validate our measurements. In this sample, which we denote as Py/Pt, a majority of in-plane current flows through the top Pt layer ($f_{Pt} \approx 70\%$); the bottom Ta layer with high resistivity accommodates only ≈10% of the total current [21]. It has also been shown that the total current-induced field in Py/Pt lies along the direction of $H_{Oe}$ [18,21].

To quantify the in-plane current-induced transverse field, we employed the second-order PHE technique (Fig. 1), originally developed by Fan *et al*. [18,19]. For Py thin films, the PHE signal from in-plane magnetization tilting dominates over any anomalous Hall effect (AHE) signal from out-of-plane tilting [18]. As such, the second-order PHE voltage $\Delta V_{PH} = V_{PH}(+I_{dc})+V_{PH}(-I_{dc})$, with $V_{PH} = V_{+}-V_{-}$ in Fig. 1(a), is related to the in-plane magnetization component transverse to the current axis. The second-order PHE is thus sensitive to small magnetization tilting induced by the total current-induced transverse field $H_{I,tot}$, i.e., the sum of the Oersted field $H_{Oe}$ and spin-orbit field $H_{so}$, as illustrated in Fig. 1(a).

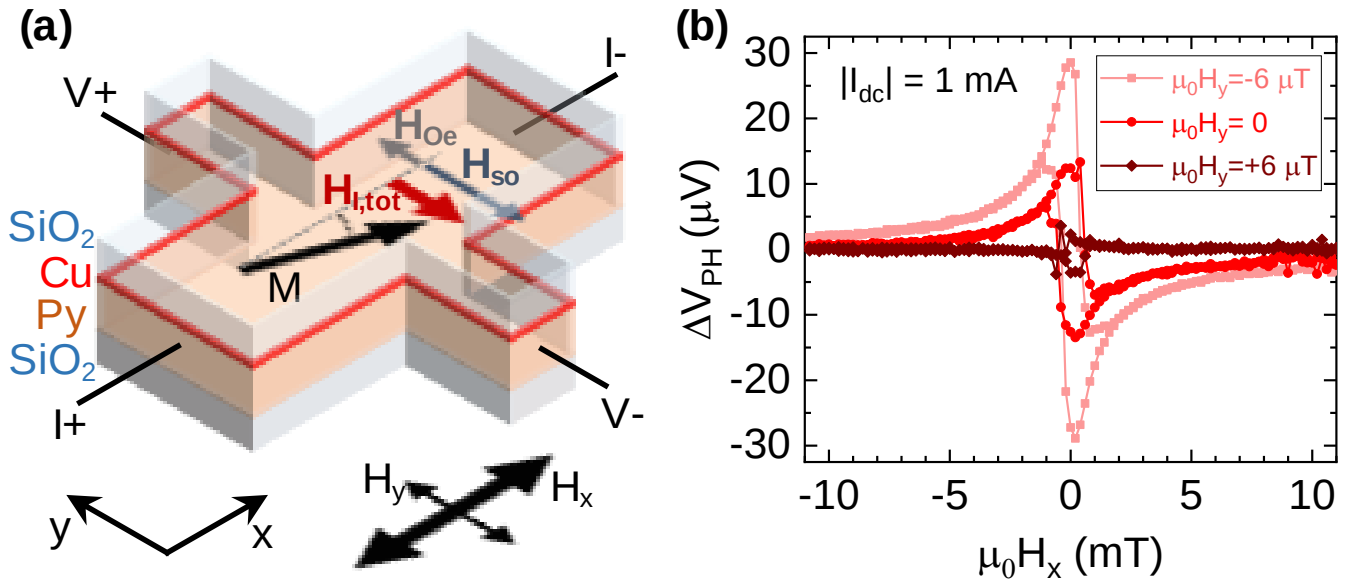

**Figure 1.** (a) Schematic of the second-order PHE measurement. Here, the total current-induced field $H_{I,tot}$ (dominated by a sizable spin-orbit field $H_{so}$) opposes the Oersted field $H_{Oe}$. Note that $H_{I,tot} = H_{so} + H_{Oe}$. (b) Example second-order PHE curves for a 100-µm-wide Py/Cu sample, obtained at $|I_{dc}| = 1$ mA.

We obtained $H_{I,tot}$ directly from the in-plane transverse calibration field $H_y$ that nulls the second-order PHE voltage. Figure 1(b) shows exemplary second-order PHE results at a drive current of $|I_{dc}| = 1$ mA in 100-µm-wide Py/Cu, measured with a probe station inside a two-axis Helmholtz coil setup. When a finite transverse calibration field $H_y$ is applied, the second-order PHE voltage is expressed as $\Delta V_{PH} = V_{PH}(+I_{dc}, +H_y)+V_{PH}(-I_{dc}, -H_y)$ [18,19]. In Fig. 1(b), $\mu_0|H_y| \approx 6$ µT along $+y$ nulls the PHE voltage, which signifies that 1 mA in the $+x$-direction generates $\mu_0|H_{I,tot}| \approx 6$ µT in the $-y$ direction. Our measurements near this nulled limit (e.g., $\mu_0 H_y = +6$ µT in Fig. 1(b)) show that the second-order Hall voltage converges to zero at large positive and negative swept fields $H_x$. This observation confirms the absence of any significant AHE [18] or thermoelectric contributions (e.g., spin Seebeck and anomalous Nernst effects) [22] that would produce a sizable difference in the saturated Hall voltages at large positive and negative $H_x$. For results shown in the remainder of this Letter, we used transverse calibration fields $\mu_0 H_y = +100$ µT and $-100$ µT and extrapolated $H_{I,tot}$, as previously used in Refs. [12,14,19] and summarized in the Supplementary Material. We note that in Py/Cu, the observed $H_{I,tot}$ lies *opposite* to $H_{Oe}$ (Fig. 1), suggesting the presence of a sizable spin-orbit field $H_{so}$ (Eq. 1) as further discussed later in this Letter.

The total current-induced transverse field $H_{I,tot}$ obtained with the second-order PHE technique is summarized in Fig. 2. In sym-Py, $H_{I,tot}$ is negligible as expected from the nominally symmetric current distribution. By contrast, $H_{I,tot}$ increases linearly with driving current $|I_{dc}|$ for Py/Ti, Py/Cu, and Py/Pt. One contribution to the observed $H_{I,tot}$ is the Oersted field $H_{Oe}$, which arises due to the higher current distribution in the top portion of the stack structure. However, as noted above and shown in Fig. 2(b,c), the direction of $H_{Oe}$ is opposite to that of the observed $H_{I,tot}$ in Py/Ti and Py/Cu. We emphasize that the calculated $H_{Oe}$ (dashed line in Fig. 2) for each stack structure is the realistic upper bound: if the in-plane current is more uniformly distributed between the ultrathin metal and Py, then the magnitude of $H_{Oe}$ is smaller.

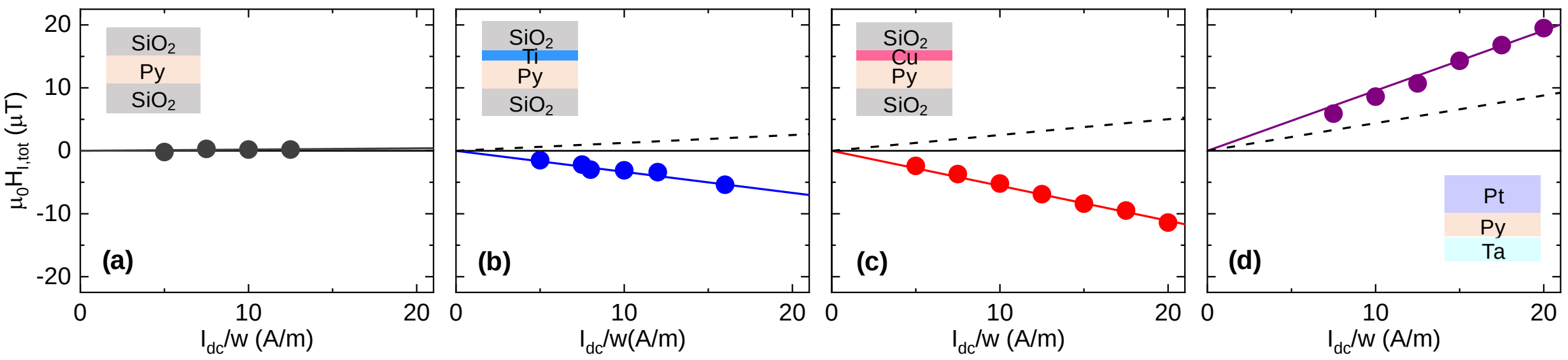

**Figure 2.** The total current-induced field $H_{I,tot}$ measured with the second-order PHE technique for (a) sym-Py, (b) Py/Ti, (c) Py/Cu, and (d) Py/Pt, plotted vs the dc current $I_{dc}$ normalized by the device width $w$ = 100 μm. The dashed lines in (b-d) indicate the estimated Oersted field. Uncertainty of the measured $H_{I,tot}$ is within the size of the dots.

Evidently, the broken symmetry with an ultrathin layer of weak spin-orbit metal (i.e., Ti or Cu) gives rise to a spin-orbit field $H_{so}$ (Eq. 1), which opposes and is at least 3 times larger than $H_{Oe}$. While a similar $H_{so}$ has been reported before [12], our present study directly shows that ultrathin insertion layers of Ti and Cu yield the same direction of $H_{so}$. This observation, in contrast to the opposite signs of the bulk spin-Hall effect in Ti and Cu [23], indicates that $H_{so}$ here is unrelated to the filling of *d*-orbitals in Ti and Cu.

The Py/Pt control sample validates our second-order PHE results. The observed $H_{I,tot}$ in Py/Pt lies in the same direction as $H_{Oe}$ (Fig. 2(d)), consistent with prior reports [18,21]. Moreover, we confirm that the magnitude of $H_{so}$ is approximately double that of $H_{Oe}$ in Py/Pt, consistent with the dc-biased ST-FMR study on the same stack structure [21]. We remark that a prior experimental study [12] shows *suppression* of $H_{so}$ when Py is interfaced with 0.5-nm-thick Pt; the origin of the significant $H_{so}$ in Py with thicker Pt (or the absence of $H_{so}$ with ultrathin Pt) is unclear and will be the subject of a future investigation.

To gain additional insight into the effects produced by in-plane current, we discuss ST-FMR results (Fig. 3) on Py/Ti, Py/Cu, and Py/Pt. While the dc-biased ST-FMR technique [12,20,21,24] enables straightforward quantitative analysis of the current-induced field (and damping-like SOT), our ST-FMR setup did not yield a sufficient signal-to-noise ratio for reliable measurement of resonance field vs dc current. Nevertheless, the ST-FMR spectral shape can qualitatively reveal the types of SOTs present (or absent) in the stack structures [18,20] as discussed in the following.

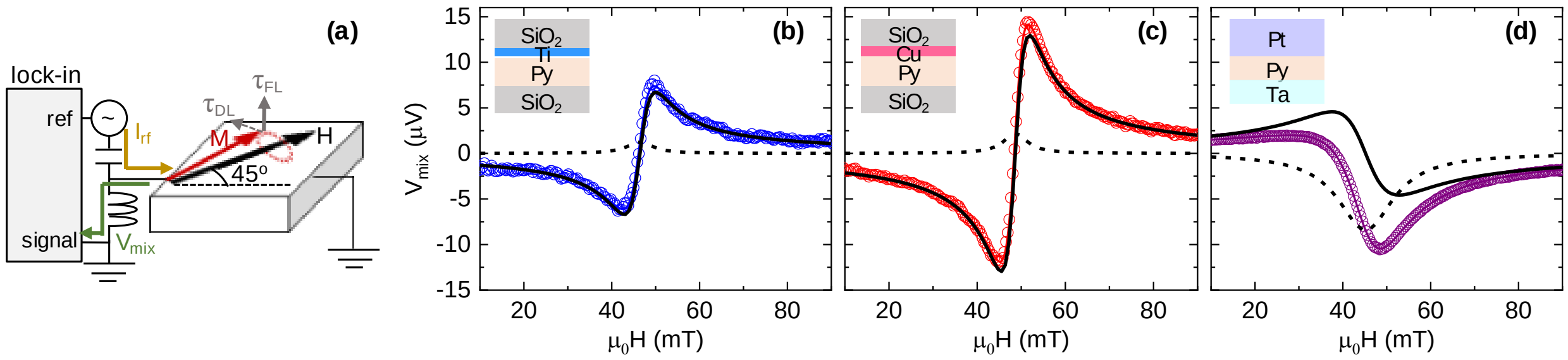

**Figure 3.** (a) Schematic of the ST-FMR measurement, driven by rf current $I_{rf}$ and detected via rectified dc voltage $V_{mix}$. (b-d) ST-FMR spectra at 5.5 GHz, +13 dBm microwave current excitation for (b) Py/Ti, (c) Py/Cu, and (d) Py/Pt. For each spectrum, the black solid curve indicates the antisymmetric component of the Lorentzian spectral fit, whereas the dashed curve indicates the symmetric component of the fit.

Figure 3(b-d) shows example ST-FMR spectra for Py/Ti, Py/Cu, and Py/Pt, each fit with a combination of antisymmetric Lorentzian (solid black curve) and symmetric Lorentzian (dashed black curve). The antisymmetric component is related to the direction of the total current-induced field [20]. The observation that Py/Ti and Py/Cu both show a large antisymmetric component opposing that of Py/Pt confirms our second-order PHE results, i.e., there is a substantial $H_{so}$ opposing $H_{Oe}$ in Py/Ti and Py/Cu. We also observe that, while Py/Pt shows a large symmetric component, Py/Ti and Py/Cu exhibit a symmetric component about an order of magnitude smaller than the antisymmetric component. This suggests that the damping-like SOT, often related to a pronounced symmetric ST-FMR spectral component [10,11,20], is negligibly

small in Py/Ti and Py/Cu compared to Py/Pt. Although identifying the origin of the small symmetric component in the ST-FMR spectra of Py/Ti and Py/Cu is beyond the scope of this Letter, it is *not* due to a damping-like SOT from partial oxidation of Cu, which would yield the same polarity of symmetric Lorentzian as Py/Pt [10].

We now discuss possible mechanisms responsible for the sizable spin-orbit field in Py/Ti and Py/Cu, as illustrated in Fig. 4. One candidate mechanism is the REE at metal-oxide interfaces (Fig. 4(a)) [12,17]. We first consider the top Py-(Ti, Cu)-$SiO_2$ interface; we lump Py/(Cu,Ti) and (Cu,Ti)/$SiO_2$ into one interface, given that the (Ti, Cu) insertion layer is only 0.5 nm thick. For both Py/Ti and Py/Cu, the spin-orbit field normalized by the estimated current density in Ti or Cu, $J_{(Ti,Cu)} = f_{(Ti,Cu)} I_{dc}/(wt)$ with $t$ = 0.5 nm, is $\mu_0 H_{so}/J_{(Ti,Cu)} \approx 0.1$ mT per $10^{11}$ A/m$^2$ This implies essentially the same magnitude of the REE for ultrathin Ti and Cu sandwiched by Py and $SiO_2$. We can estimate the Rashba coefficient $\alpha_R$ from $H_{so}/J_{(Ti,Cu)}$ through $\alpha_R \approx (\mu_B M_s/P)\mu_0 H_{so}/J_{(Ti,Cu)}$ [16,25], where $\mu_B$ is the Bohr magneton, $M_s \approx 700$ kA/m is the saturation magnetization of Py, and $P \approx 0.15$ is the current spin polarization (related to the strength of *s*-*d* exchange coupling [16]) in 3-nm-thick Py [26]. Our estimate of $\alpha_R \approx 0.003$ eV Å is an order of magnitude smaller than $\alpha_R$ from angle-resolved photoemission studies of crystalline Cu surfaces [27–29]. We remark that the interfaces of sputtered layers in our study are likely diffuse. The smallness of the estimated Rashba coefficient in our study may be due to the ill-defined interfaces of our stack structures, such that the Rashba-Edelstein field-like SOT may be enhanced with the use of highly crystalline ultrathin Ti or Cu.

The bottom $SiO_2$-Py interface might also exhibit a REE, similar to the previous claim of a REE at $Al_2O_3$-Py [12]. However, considering that Ref. [12] shows a significant spin-orbit field even in Py sandwiched between Ti and Cu, i.e., without a direct oxide-Py interface, it appears unlikely that the $SiO_2$-Py interface is the sole or dominant source. We therefore deduce that the REE at the Py-(Ti, Cu)-$SiO_2$ interface (Fig. 4(a)) dominates over that at the $SiO_2$-Py interface.

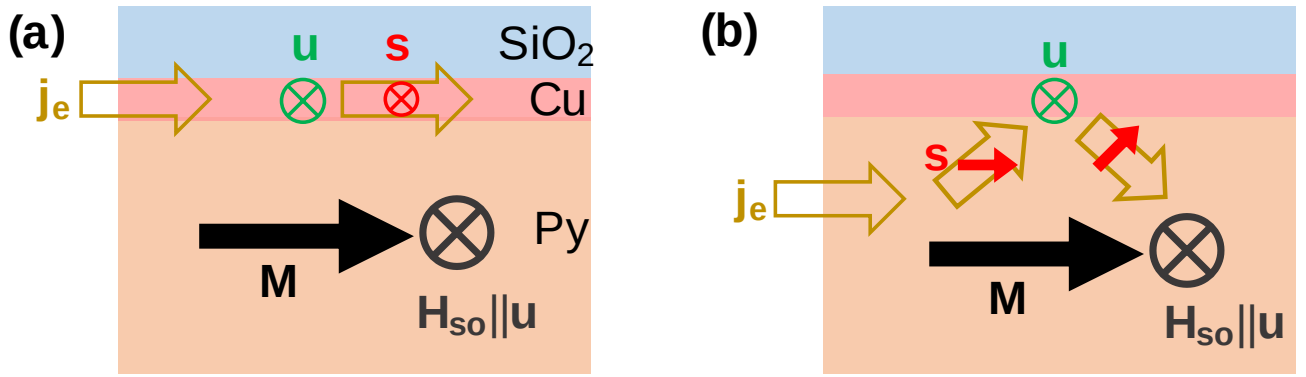

**Figure 4.** Possible mechanisms of the current-induced spin-orbit field $\mathbf{H}_{so}$ (which acts on the Py magnetization **M**) due to the interfacial Rashba field **u**. The red symbol (in (a)) and arrows (in (b)) represent the spin polarization **s** of the electron current $\mathbf{j}_e$. (a) Rashba-Edelstein effect, where the electron current flowing parallel to the Py-Cu-$SiO_2$ interface becomes spin-polarized along **u** and exchange-couples to **M**. (b) Spin-orbit precession effect, where spin-polarized conduction electrons in Py precess about **u** during reflection from the Py-Cu-$SiO_2$ interface and then exert a torque (corresponding effective field $\mathbf{H}_{so}$) on **M**.

In the REE mechanism discussed above and illustrated in Fig. 4(a), the electron current $\mathbf{j}_e$ in a quasi-two-dimensional conductor is spin-polarized by the interfacial Rashba field $\mathbf{u} \sim \mathbf{z} \times \mathbf{j}_e$, where **z** is normal to the interface; the spin-polarized electrons then generate an effective spin-orbit field $\mathbf{H}_{so}$ on the magnetization via *s-d* exchange coupling [16,17,25]. However, in our study with a 3-nm-thick conductive ferromagnet, electronic transport is actually three-dimensional. In this regard, we consider an alternative mechanism [30,31], which is illustrated in Fig. 4(b) and proceeds as follows: (1) Some conduction electrons in Py are first spin-polarized along the magnetization **M**. (2) When these polarized electrons are reflected from the Py-(Ti,Cu)-$SiO_2$ interface with the Rashba field **u**, the spin polarization precesses (rotates) about **u** and develops a finite component along $\mathbf{u} \times \mathbf{M}$ [30,31]. (3) The rotated spin polarization then dephases in Py (i.e., ultimately aligning with **M** [32]) to exert a spin torque $\boldsymbol{\tau} \sim \mathbf{M} \times \mathbf{H}_{so} \sim \mathbf{M} \times [\mathbf{M} \times (\mathbf{u} \times \mathbf{M})]$, where $\mathbf{M} \times (\mathbf{u} \times \mathbf{M}) = \mathbf{u}$. Thus, the measured spin-orbit field $\mathbf{H}_{so}$ in the Py layer points along **u**, irrespective of the magnetization direction. In other words, three-dimensional spin transport in Py – in concert with the interfacial Rashba field – may give rise to a magnetization-*independent* spin-orbit field in the ferromagnet (Fig. 4(b)) that is consistent with our experimental observations.

In summary, we have investigated the current-induced spin-orbit field (field-like SOT) in $SiO_2$-sandwiched Py, with the top Py-$SiO_2$ interface incorporating an ultrathin layer of weak spin-orbit metal, Ti or Cu. In both $SiO_2$/Py/Ti/$SiO_2$ and $SiO_2$/Py/Cu/$SiO_2$, we observe a sizable spin-orbit field opposing the Oersted field, whereas no significant damping-like SOT is found. We deduce that this spin-orbit field arises from an interfacial Rashba-Edelstein effect or spin-orbit precession primarily at the Py-(Ti, Cu)-$SiO_2$ interface. Our findings provide further insight for engineering SOTs in ferromagnets interfaced with weak spin-orbit materials.

Acknowledgements

This research was funded in part by 4-VA, a collaborative partnership for advancing the Commonwealth of Virginia, as well as by the ICTAS Junior Faculty Award. D.A.S. acknowledges support by the Virginia Tech Graduate School Doctoral Assistantship. S.E. thanks Xin Fan and Vivek Amin for helpful discussions.

**Current-induced spin-orbit field in permalloy interfaced with ultrathin Ti and Cu**

Ryan W. Greening[1], David A. Smith[1], Youngmin Lim[1], Zijian Jiang[1], Jesse Barber[1], Steven Dail[1,2],

Jean J. Heremans[1], Satoru Emori[1]

1. Department of Physics, Virginia Tech, Blacksburg, VA 24061

2. Academy of Integrated Science, Virginia Tech, Blacksburg, VA 24061

## I. Static Magnetic Properties

We characterized the static magnetic properties of the stack structures by performing vibrating sample magnetometry (Microsense EZ9) on mm-scale squares from the same wafers as the patterned Hall crosses and ST-FMR microstrips. The saturation magnetization of ≈600 kA/m for sym-Py is slightly lower than ≈700 kA/m for Py/Ti and Py/Cu, possibly because of partial oxidation of the top surface of Py in direct contact with $SiO_2$. We also find that the coercivity of sym-Py (≈0.8 mT) exceeds that of Py/Ti (≈0.5 mT) and Py/Cu (≈0.3 mT). The enhanced coercivity for sym-Py is consistent with the presence of an ultrathin antiferromagnetic oxide layer at the Py-$SiO_2$ interface (e.g., NiO) that is exchange-coupled to Py [S1].

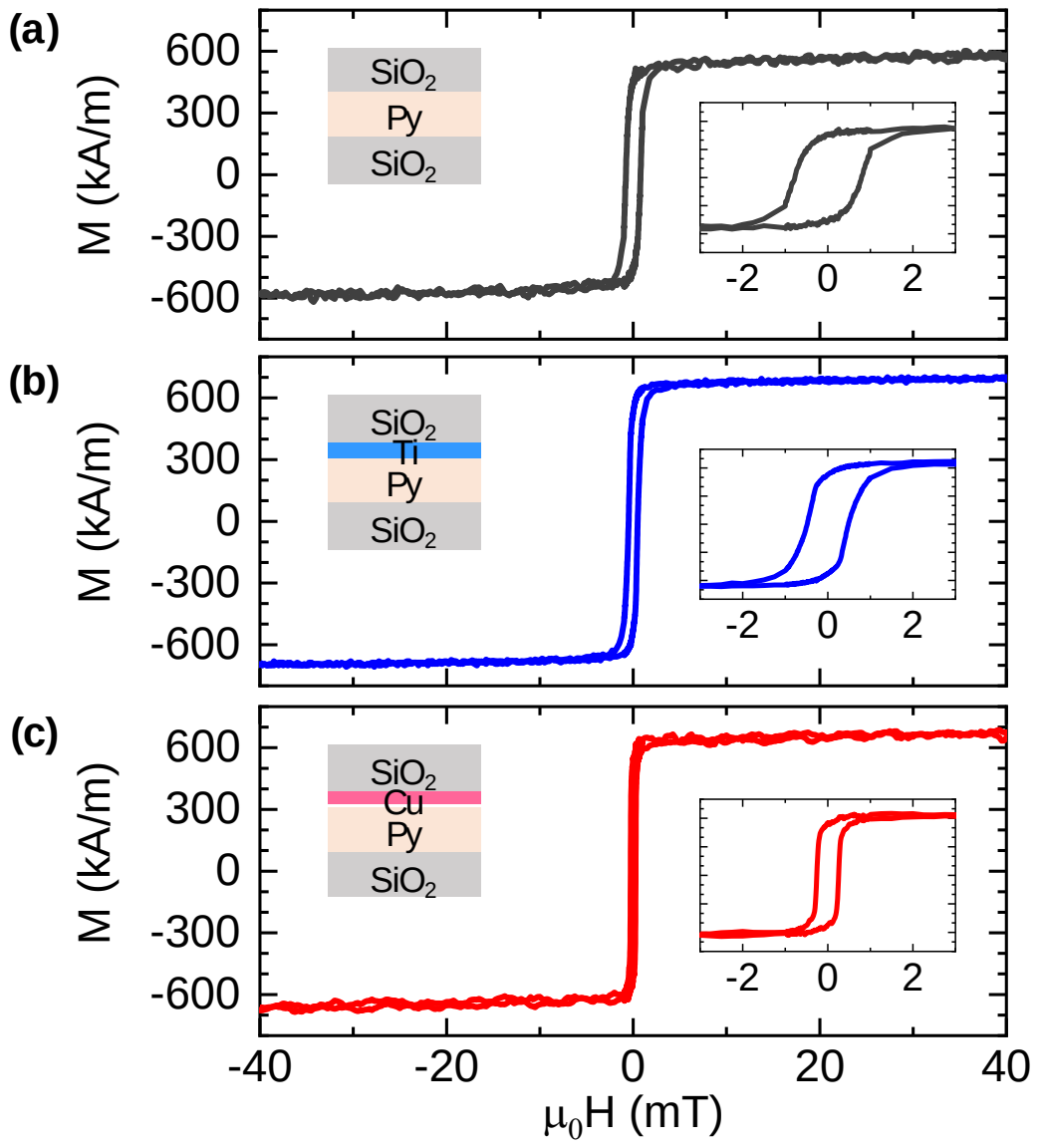

Figure S1. Vibrating sample magnetometry with *H* applied in the film plane.

## II. Extraction of the Total Current-Induced Transverse Field $H_{I,tot}$: Extrapolation Method

Here, we summarize the extrapolation method (similar to that used in Refs. [S2–S4]) used to extract the total current-induced transverse field $H_{I,tot}$. For a uniform magnetization with a small deviation from the current axis ($x$-axis in Figure 1 of the main text), the second-order planar Hall effect (PHE) voltage $\Delta V_{PH}$ is proportional to the $y$-component of the magnetization $\Delta m_y$, and hence the sum of $H_{I,tot}$ and transverse calibration field $H_y$ [S4,S5],

$$\Delta V_{PH} \propto \Delta m_y \propto H_{I,tot} + H_y.$$

It follows that $\Delta V_{PH}$ at $H_y = 0$ is expressed as

$$\Delta V_{PH}(H_y = 0) \propto H_{I,tot}.$$

The *difference* of $\Delta V_{PH}$ at a fixed $\mu_0|H_y| = 100$ µT, which we call $\Delta V_{fit}$, is

$$\Delta V_{fit}(\mu_0|H_y| = 100\ \mu\text{T}) = \Delta V_{PH}(\mu_0 H_y = +100\ \mu\text{T}) - \Delta V_{PH}(\mu_0 H_y = -100\ \mu\text{T}) \propto 2H_y.$$

Thus, by plotting $\Delta V_{PH}$ versus $\Delta V_{fit}$, we can quantify $H_{I,tot}$ from

$$\frac{\Delta V_{PH}}{\Delta V_{fit}} = \frac{\mu_0 H_{I,tot}}{200\ \mu\text{T}}.$$

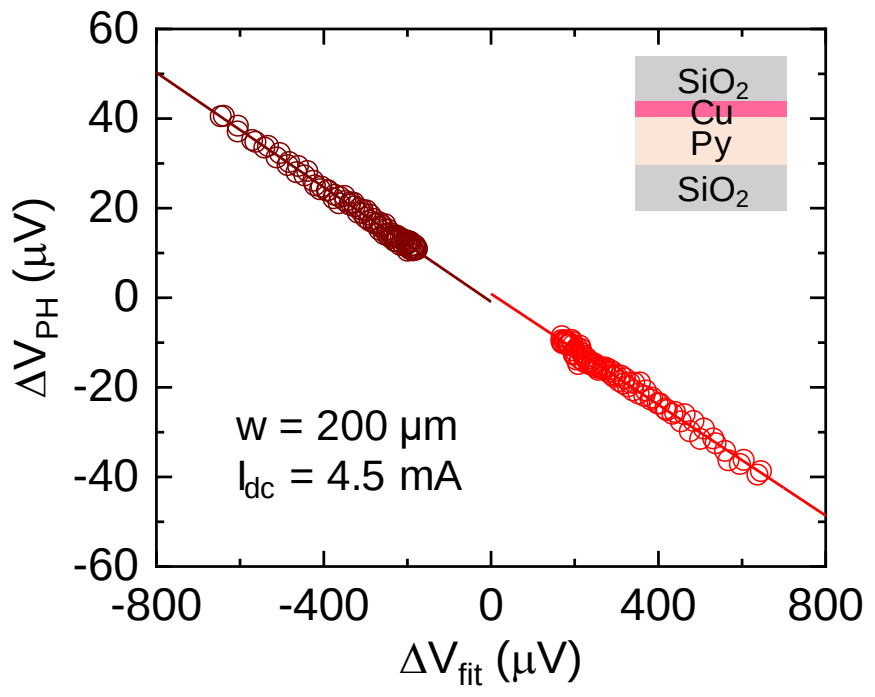

Figure S2. Example linear fitting in the PHE extrapolation method. Here, $\mu_0 H_{I,tot}$ = –13 µT at $I_{dc}/w$ = 22.5 A/m.

Figure S2 shows an example of the extrapolation method. For the above-described linear fitting in our study, we only use data obtained at sufficiently large external magnetic fields (i.e., $\mu_0|H_x| > 2.5$ mT) to ensure that the magnetization is uniform with a small-angle deviation from the $x$-axis.

In this extrapolation method, any anomalous Hall effect (AHE) or thermoelectric contributions in the second-order Hall voltage would appear as a significant vertical offset (discontinuity) between the separate linear fits for $\Delta V_{fit} > 0$ and $\Delta V_{fit} < 0$. Such an offset was observed in a previous study by Fan *et al*. (e.g., Figure 2(c) of Ref. [S4]). By contrast, no significant offsets are observed for samples in our study (e.g., Figure S2). We therefore conclude that AHE and thermoelectric contributions to the second-order Hall voltage are negligible in our study.